\documentclass[showpacs,amsmath,amssymb,aps,twocolumn,prl]{revtex4-1}
\usepackage{graphicx}
\usepackage{soul}
\usepackage[colorlinks=true,citecolor=blue,linkcolor=magenta]{hyperref}
\usepackage[usenames]{color}
\usepackage[cspex,bbgreekl]{mathbbol}
\usepackage{relsize}
\usepackage{float}

\newcommand {\grsim} {\ {\raise-.5ex\hbox{$\buildrel>\over\sim$}}\ }
\newcommand {\lessim} {\ {\raise-.5ex\hbox{$\buildrel<\over\sim$}}\ }
\newcommand {\ii} {i}
\newcommand {\si} {Supplementary Information}

\begin{document}

\title{Observation of the Meissner effect with ultracold atoms in bosonic ladders}

\author{M. Atala$^{1,2}$, M. Aidelsburger$^{1,2}$, M. Lohse$^{1,2}$, J. T. Barreiro$^{1,2}$, B. Paredes$^{3}$ \& I. Bloch$^{1,2} $}

\affiliation{$^{1}$\,Fakult\"at f\"ur Physik, Ludwig-Maximilians-Universit\"at, Schellingstrasse 4, 80799 M\"unchen, Germany\\
$^{2}$\,Max-Planck-Institut f\"ur Quantenoptik, Hans-Kopfermann-Strasse 1, 85748 Garching, Germany\\
$^{3}$\,Instituto de F\'{i}sica Te\'{o}rica CSIC/UAM \\C/Nicol\'{a}s Cabrera, 13-15
Cantoblanco, 28049 Madrid, Spain}
\date{\today}


\begin{abstract}
We report on the observation of the Meissner effect in bosonic flux ladders of ultracold atoms. Using artificial gauge fields induced by laser-assisted tunneling, we realize arrays of decoupled ladder systems that are exposed to a uniform magnetic field. By suddenly decoupling the ladders and projecting into isolated double wells, we are able to measure the currents on each side of the ladder. For large coupling strengths along the rungs of the ladder, we find a saturated maximum chiral current corresponding to a full screening of the artificial magnetic field. For lower coupling strengths, the chiral current decreases in good agreement with expectations of a vortex lattice phase. Our work marks the first realization of a low-dimensional Meissner effect and, furthermore, it opens the path to exploring interacting particles in low dimensions exposed to a uniform magnetic field.\end{abstract}

\maketitle
The Meissner effect is the hallmark signature of a superconductor exposed to a magnetic field \cite{Meissner:1933,Bardeen:1957}. For a type-II superconductor, full screening of the applied external field occurs up to a critical field $H_{c1}$. Such a screening is the result of circular surface currents on the superconductor that generate an opposite field, canceling the applied field. The superconductor thus acts as a perfect diamagnet in the Meissner phase. For larger field strengths $H>H_{c1}$, however, the superconductor is not able to fully screen the applied field and an Abrikosov vortex lattice phase is formed in the system. In low-dimensional quantum systems it has been a longstanding challenge to probe analogue ideas and to investigate the interplay of orbital magnetic field effects and interactions. While a single one-dimensional system does not allow for any orbital magnetic field effects, a ladder system is the simplest extension where these are permitted \cite{Orignac:2001,Petrescu:2013,Dhar:2012,Huegel:2013,Celi:2013,Kessler:2013}.

\begin{figure}[t!]
\includegraphics{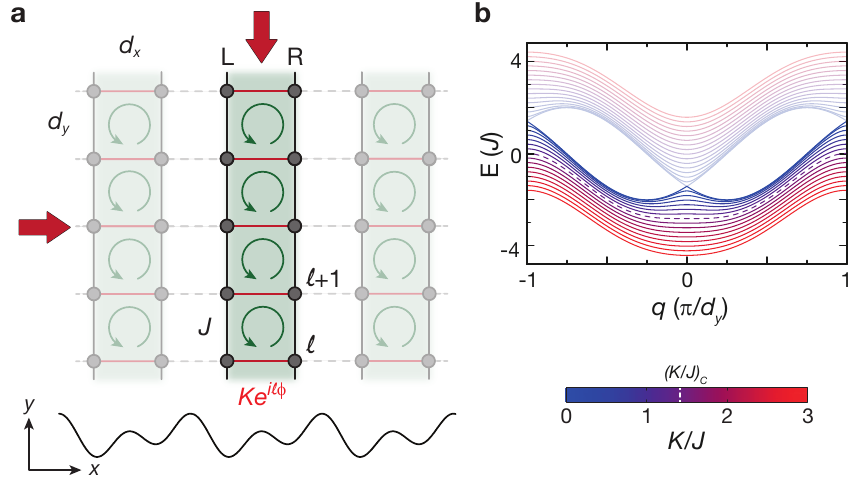}
\vspace{-0.cm} \caption{Experimental setup and energy bands. \textbf{(a)} The experiment consists of a one-dimensional array of ladders in the $xy$-plane with lattice constants given by $d_{x,y}$. An effective homogeneous magnetic field in each ladder is realized through laser-induced tunneling (red arrows) between the left (L) and right (R) legs of the ladder \cite{Aidelsburger:2011,Supplements}, which leads to spatially dependent complex tunneling amplitudes along the rungs of the ladder $Ke^{ i \ell \phi}$, with constant magnitude $K$. The flux per plaquette is then given by $\phi$, where in our experiments $\phi=\pi/2$ is realized. \textbf{(b)} Band structure of the flux ladder for different values of $K/J$ and flux $\phi=\pi/2$. The system exhibits two bands, colored with a blue-red gradient that indicates the value of $K/J$. The lowest band presents a single minimum in the Meissner phase $K/J>(K/J)_c$ and two symmetric minima in the vortex phase $K/J<(K/J)_c$, where $(K/J)_c=\sqrt{2}$. In the spin-orbit representation, the legs of the ladder play the role of pseudo-spins.\label{Fig_1}}
\end{figure}

Here we report on the realization of such bosonic ladders for ultracold atoms exposed to a uniform artificial magnetic field created by laser-assisted tunneling \cite{Goldman:2013,Jaksch:2003,Gerbier:2010,Kolovsky:2011,Aidelsburger:2011,Aidelsburger:2013,JimenezGarcia:2012,Struck:2012, Miyake:2013}. Previously, such ladders have been discussed in the context of Josephson-junction arrays \cite{Kardar:1986,Granato:1990,Denniston:1995,Nishiyama:2000,Orignac:2001} and more recently also for ultracold atoms exposed to an artificial gauge field \cite{Huegel:2013,Celi:2013,Kessler:2013}. In our experiment we can measure the probability current on either leg of the ladders and, in addition, observe the momentum distribution of the system after time-of-flight expansion. Rather than varying the external field strength, we determine the response of the system as a function of the ratio of transverse rung coupling $K$ to coupling along the legs of the ladder $J$ (see Fig.~\ref{Fig_1}). In full analogy to the type-II superconductor, we find evidence for a Meissner phase with maximum chiral currents that screen the applied field. Below a critical coupling strength $(K/J)_c$ we find a decreasing chiral current, in good agreement with theoretical expectations for a vortex phase with only partial screening.\\

\begin{figure*}[t!]
\includegraphics{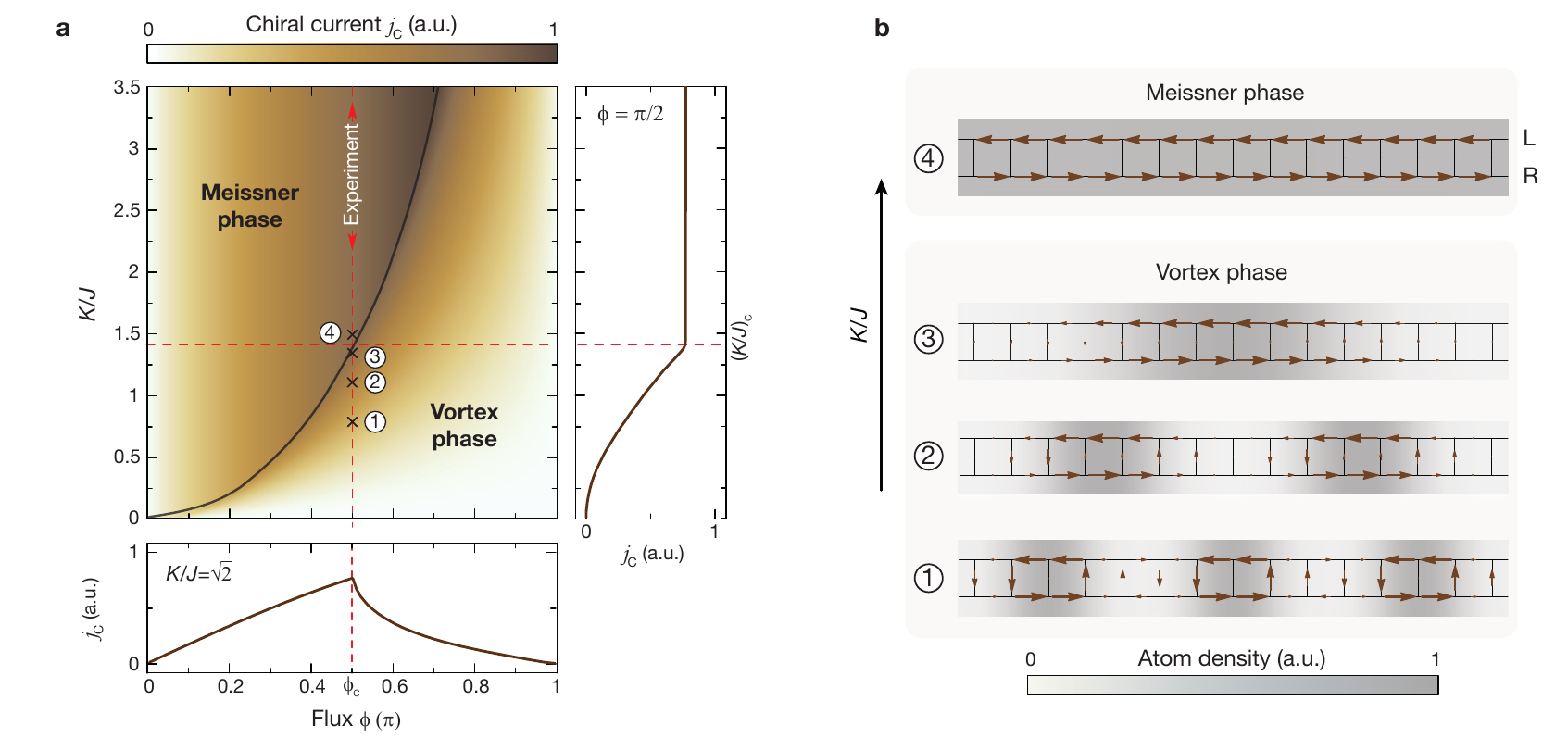}
\vspace{-0.cm} \caption{Phase diagram of ladder currents. \textbf{(a)} Chiral current strength along the legs of the ladder as a function of the flux $\phi$ and the ratio $K/J$ calculated for the ground state of a ladder with $300$ sites and periodic boundary conditions. The solid line shows the critical boundary that separates the Meissner and the vortex phase. The lower graph shows a profile of the current for a fixed value of $K/J=\sqrt{2}$. In the Meissner phase the current increases with flux, until a critical flux $\phi_c$ is reached after which the system transitions into a vortex phase where the chiral current decreases with flux. The graph on the right is a profile line for a fixed flux $\phi=\pi/2$, as used in the experiment. In that case, the chiral current increases in the vortex phase when increasing $K/J$ until one enters the Meissner phase at the critical point $(K/J)_c$ where the current saturates. \textbf{(b)}  Theoretically calculated individual currents and particle densities for the different values of $K/J$ marked in (a). The thickness and length of the arrows denotes the current strength, which is normalized to the maximum current for each $K/J$. To clearly illustrate the modulation of the density, we have subtracted a homogeneous offset and renormalized the density modulation for each $K/J$.
\label{Fig_2}}
\end{figure*}

The simplest theoretical description for our system is that of non-interacting bosonic particles in an infinitely extended two-leg ladder geometry, subject to a magnetic flux $\phi$ per plaquette. The corresponding Hamiltonian is:
\begin{align}
H = &-J  \underset{\ell}{\sum}\left(\hat a^{\dagger}_{\ell+1;L}\hat a^{}_{\ell;L}+\hat a^{\dagger}_{\ell+1;R}\hat a^{}_{\ell ;R} \right) \nonumber \\
& -K\underset{\ell}{\sum} \left ( e^{-i\ell \phi}\hat a^{\dagger}_{\ell;L}\hat a^{}_{\ell;R}\; \right )+ \text{h.c.}  \label{eq:H}
\end{align}
Here, the operator $\hat a^{}_{\ell;\mu}$ annihilates a particle at site $\ell$ in the left or right leg of the ladder, where $\mu=(L,R)$. The hopping amplitude between neighbouring sites along the ladder is $J$, and $K e^{i \ell\phi}$ denotes the spatially dependent tunneling amplitude between legs. This Hamiltonian can be mapped onto a spin-orbit coupled system, where the pseudo-spin represents the legs of the ladder \cite{Huegel:2013,Goldman:2013}. Observables that can be readily measured in the experiment, and that allow one to characterize the different phases of the system, are the gauge-independent average current on either side of the ladder $j_\mu=N^{-1}_{leg} \sum_{\ell} \langle \hat j_{\ell,\ell+1;\mu}\rangle$ and the chiral current $j_C = j_L-j_R$ \cite{Huegel:2013}. Here $N_{leg}$ is the number of sites along the ladder and $\hat j_{\ell,\ell+1;\mu}$ denotes the current operator for currents flowing from site $\ell \rightarrow \ell+1$.


For low flux values $\phi \leq \phi_c$, the ground state of the Hamiltonian exhibits a Meissner phase (see Fig.~\ref{Fig_2}), with maximal and opposite currents along the two legs of the ladder $|j_\mu|=(2J/\hbar) \sin(\phi/2)$, i.e., a full screening of the applied magnetic field. Increasing the flux leads to increasing edge currents up to a critical flux $\phi_c$ beyond which the current abruptly starts to decrease. At this point the system enters a vortex phase with decreasing edge currents, where the magnetic field partially penetrates the system. Such a behaviour exactly parallels the one of the Meissner effect in a type-II superconductor and its transition into an Abrikosov vortex lattice phase. For a neutral superfluid in a thin rotating annulus, the Hess-Fairbank effect has been discussed as an analogue of the Meissner effect \cite{Hess:1967,Ramanathan:2011}. The phase transition from the Meissner to the vortex phase in our system is characterized by a change in the band structure, where the single minimum at $q=0$ in the lower band splits into two minima at finite $q$ (Fig.~\ref{Fig_1}b). In our experiment we chose the following strategy to observe the transition from a Meissner to a vortex phase: rather than changing the magnetic field strength, we worked at a fixed flux and varied the rung-to-leg coupling ratio $K/J$. As can be seen in Fig.~\ref{Fig_2}a, in this case one expects to observe an increase in the leg currents for increasing $K/J$ up to a critical coupling strength $(K/J)_c$ after which a saturation in the current occurs, signaling the transition from the vortex to the Meissner phase. In the vortex phase, the wave function exhibits a vortex structure for the currents, whose period increases with $K/J$, and the atom density on the ladder also becomes modulated with the same periodicity. In the Meissner phase, on the other hand, the size of the vortex is infinite and the density is uniform (Fig.~\ref{Fig_2}b).

Our experimental setup consists of a Bose-Einstein condensate of $^{87}$Rb atoms loaded into a three dimensional optical lattice potential. This potential is created by a standing wave of wavelength $\lambda_s=767$\,nm along $y$ and a superposition of a short and a long standing wave of wavelengths $\lambda_s$ and $\lambda_{l}=2 \lambda_s$ respectively along $x$. Additionally, a weak standing wave of $\lambda_z=844$\,nm that does not isolate different planes is used along $z$. The resulting superlattice potential in the $x$ direction is of the form $V(x)=V_{lx}\textrm{sin}^2(k_{l}x+\varphi /2)+V_{x}\textrm{sin}^2(k_{s}x)$, where $k_{i}=2\pi/\lambda _{i}$, $i\in \{s,l\}$. The lattice depths $V_{lx/x}$ and phase $\varphi$ were chosen to have an array of isolated tilted double well potentials along $x$, where each double well corresponds to a single realization of a ladder. Using the same scheme as in our previous works \cite{Aidelsburger:2011,Aidelsburger:2013}, we employ a pair of far-detuned running-wave beams to induce left-right tunneling inside each double well. This lattice configuration creates a one-dimensional array of isolated ladders in the $xy$-plane, with a total flux per plaquette $\phi=\pi/2$ (see Fig.~\ref{Fig_1}a and \cite{Supplements}).

\begin{figure}
\includegraphics{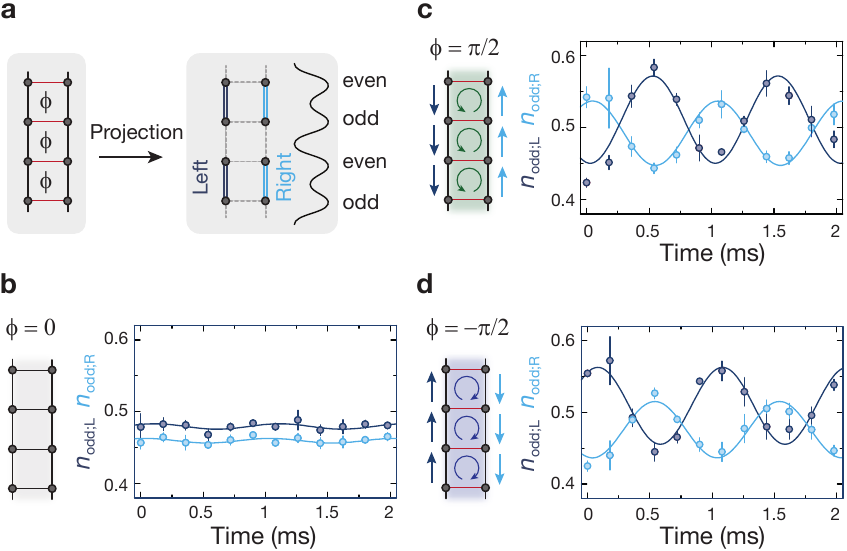}
\vspace{-0.cm} \caption{Current measurement. \textbf{(a)} Schematics of the projection into isolated double wells along $y$ used to measure the currents. Current measurements are carried out simultaneously for every other bond. \textbf{(b-d)}  Time evolution of the population fractions $n_{\textrm{odd;L}}(t)$ (dark blue) and $n_{\textrm{odd;R}}(t)$ (light blue) for the fluxes $\phi=0$ (b), $\phi=+\pi/2$ (c) and $\phi=-\pi/2$ (d). The expected current in the case $\phi=0$ is zero and therefore there are no even-odd oscillations. When the flux $\phi=+\pi/2$ ($\phi=-\pi/2$) the edge current flows counterclockwise (clockwise). For (c, d) the coupling ratio was $K/J=330(20)$\,Hz$/150(10)$\,Hz$\,=2.2(1)$. The small initial phase offsets of $n_{\textrm{odd};\mu}(t)$ are caused by the finite projection time. Each experimental point is an average over three measured values and the error bars depict the corresponding standard deviations. The solid lines are sinusoidal fits to the experimental data, where the frequency was fixed to the calibrated double well coupling $2J_D/\hbar$. The atom fractions $n_{\textrm{odd};\mu}(t)$ and $n_{\textrm{even};\mu}(t)$ were determined by transferring the atoms to higher lying Bloch bands and applying a subsequent band mapping technique~\cite{Supplements,SebbyStrabley:2006}. \label{Fig_3}}
\end{figure}

In order to reveal the presence of the chiral edge currents, we prepared the system in the ground state of the flux ladder and measured the currents on the left and right legs averaged over an array of about 15 individual ladders. In the experimental sequence we loaded a Bose-Einstein condensate of about $5\times10^4$ atoms into the isolated flux ladders for different values of $K/J$ (see~\cite{Supplements} for a description of the experimental sequence).  To extract the currents along the legs, the wave function was then suddenly projected into isolated double wells along $y$ and held for a certain holding time $t$ in this configuration (see Fig.~\ref{Fig_3}a and ~\cite{Supplements,Trotzky:2012,Kessler:2013}). During the projection we also decoupled the legs of the ladder by switching off the left-right laser-assisted tunneling such that atoms can only tunnel within a single double well along $y$.  If $J_D$ is the tunnel coupling inside each individual double well, then the averaged even-odd atom fraction oscillates according to
\begin{align}
n_{\textrm{even};\mu}(t)-n_{\textrm{odd};\mu}&(t) = \nonumber \\
&\left [n_{\textrm{even};\mu}(0)-n_{\textrm{odd};\mu}(0) \right ]\textrm{cos}(2J_\textrm{D}t/\hbar) \nonumber \\
&-\frac{j_{\mu}}{J/\hbar}\textrm{sin}(2J_\textrm{D}t/\hbar), \label{eq:poposc}
\end{align}

\noindent where $n_{\textrm{odd};\mu}(t)=\sum_{\ell} n_{2\ell+1;\mu}(t)$ and  $n_{\textrm{even};\mu}(t)$ $=\sum_{\ell} n_{2\ell;\mu}(t)$ are the averaged atom fractions over the individual double wells, with $n_{\ell;\mu}(t)=\langle \hat a^{\dagger}_{\ell;\mu}\hat a^{}_{\ell;\mu}\rangle/\sum_{\ell} \langle a^{\dagger}_{\ell;\mu}\hat a^{}_{\ell;\mu}\rangle$. The quantity $j_\mu=N^{-1}_{leg} \sum_{\ell} \langle \hat j_{2\ell,2\ell+1;\mu}\rangle$ is the normalized average of the currents on the left or right side of the ladder where the expectation value is calculated for the initial state directly after the projection. Though the average runs only over the bonds within the projected double wells (i.e., every other bond, see Fig.~\ref{Fig_3}a), it is a very good approximation of the average leg currents for our system. The first oscillating term of the atom fraction in Eq.~\ref{eq:poposc} is proportional to the initial population imbalance and should be ideally zero due to the averaging. The second term is proportional to the current amplitude and dominates the time evolution. Therefore, currents with opposite directions along the legs result in population oscillations in the double wells that are out of phase by $\pi$. In Fig.~\ref{Fig_3}(b-d) the experimentally measured time evolution $n_{\textrm{odd;}\mu}(t)$ for positive, negative and zero flux are displayed. For $\phi=+\pi/2$ the current flows downwards on the left leg and upwards on the right leg yielding an even-odd oscillation of $n_{\textrm{odd;L}}(t)$ and $n_{\textrm{odd;R}}(t)$ with an initial phase of $\pi$ and zero, respectively. When the flux is reversed to $\phi=-\pi/2$, the phases of $n_{\textrm{odd;L}}(t)$  and $n_{\textrm{odd;R}}(t)$ are also reversed. This demonstrates that the flux ladder exhibits a chiral edge current in the ground state whose chirality is reversed when inverting the direction of the flux, in agreement with the theoretical expectation. For the case without the applied artificial magnetic field ($\phi=0$), the wave function is homogeneous throughout the ladder and no chiral currents are present, leading to a vanishing oscillation amplitude in the double wells, as observed in the experiment (see \cite{Supplements} for a detailed description of the experimental sequence).

\begin{figure}
\includegraphics{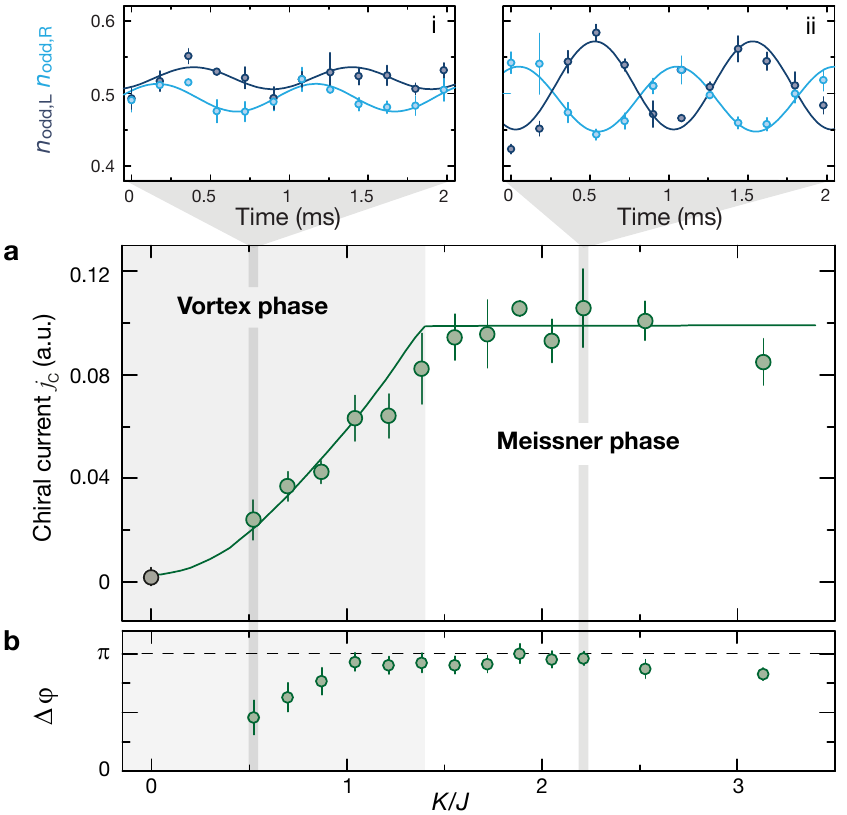}
\vspace{-0.cm} \caption{Chiral current as a function of $K/J$. \textbf{(a)} Average chiral current $j_C$ obtained from sinusoidal fits to the individual oscillations of $n_{\textrm{odd;}\mu}(t)$ for different values of $K/J$. The darker (lighter) area indicates the vortex (Meissner) phase. The solid green line is a theory curve fitted to the experimental data, which was calculated using the Hamiltonian of Eq.~\ref{eq:H}. The fitted amplitude and offset were 0.14(1) and 0.001(2) respectively \cite{Supplements}. The gray point was measured in a ladder with $\phi=0$ where the chiral current is zero. The two insets above show the average of three individual oscillations for $K/J=0.52(7)$ (i) and for $K/J=2.2(1)$ (ii).  \textbf{(b)} Phase difference $\Delta \varphi$ between the oscillations $n_{\textrm{odd;L}}(t)$ and $n_{\textrm{odd;R}}(t)$ for different values of $K/J$.  We observe that $\Delta \varphi \approx \pi$ for $K/J>1$ and decreases when $K/J \lesssim 1$. All data points were extracted from three individual measurements of $n_{\textrm{odd;}\mu}$. For the chiral current, the data points were evaluated through $|I_{\textrm{L}}-e^{\ii \Delta \varphi}I_{\textrm{R}}|$, where we averaged the three independently fitted amplitudes $I_\mu$ and calculated the resulting standard deviations. For the phase measurements we fitted the phase of the average of the three independent oscillations and from the errors of the fits we determined the error bars.\label{Fig_4}}
\end{figure}

\begin{figure}
\includegraphics{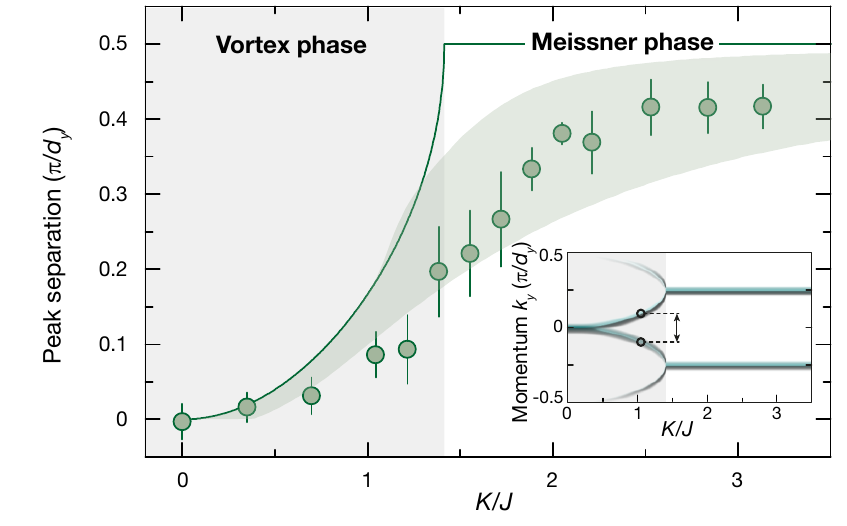}
\vspace{-0.cm} \caption{Relative position of the momentum peaks. Experimental peak separation between inner peaks as a function of $K/J$ fitted from the time-of-flight images. Each point corresponds to an average of 5-40 individual measurements and the error bars are the standard deviations. The solid line is the theoretically calculated peak separation, where there is no free parameter. The light green shaded area shows the peak separation calculated for a system with a density of 25 particles per single site of the ladder and for a temperature range from 10nK to 30nK \cite{Supplements}. The inset shows the expected momentum distribution along $y$ as a function of $K/J$, and the black circles highlight the measured peak separation.\label{Fig_5}}
\end{figure}

In order to probe the phase diagram shown in Fig.~\ref{Fig_2}a, we studied the change of the chiral current amplitude when increasing the ratio $K/J$ for a constant flux  $\phi=\pi/2$. On each leg of the ladder we measured $n_{\textrm{odd;}\mu}(t)$ and fitted its amplitude $I_\mu$ and phase $\varphi_\mu$ for different values of $K$ and constant $J$. To extract the chiral current, we made use of the left-right symmetry of the wave function density i.e., $n_{\textrm{even};L}(0)-n_{\textrm{odd};L}(0)=n_{\textrm{even};R}(0)-n_{\textrm{odd};R}(0)$, from which we obtain $j_C = j_L-j_R\propto|I_{\textrm{L}}-e^{\ii \Delta \varphi}I_{\textrm{R}}|$, with $\Delta \varphi=\varphi_R-\varphi_L$. As shown in Fig.~\ref{Fig_4}, in the vortex phase the chiral current increases when increasing $K/J$ up to the critical point $(K/J)_c=\sqrt{2}$ at which the system enters the Meissner phase indicated by a saturation of the chiral current. For a comparison with theory, we fit the theoretically predicted behaviour with amplitude and offset as free fit parameters and find good agreement between theory and experiment. We also observe that when $K/J>1$, the value of the phase difference $\Delta \varphi$ is close to $\pi$. This is to be expected whenever the averaged initial population imbalance on neighbouring sites in the double well is negligible (see Eq.~\ref{eq:poposc}) and a chiral current is present in the system. For values of $K/J<1$, we find that $\Delta \varphi$ decreases, most likely due to the fact that the smaller and smaller leg currents for decreasing $K/J$ lead to a larger effect of any possible initial population imbalance on the phase of the double well oscillations (see Eq.~\ref{eq:poposc}). These population differences are not perfectly averaged out over the entire system in the experiment and lead to the observed decrease in $\Delta \varphi$ for small values of $K/J$ (see Eq.~\ref{eq:poposc}). We note however that by subtracting the two population oscillations, as described above for the left and right leg of the ladder, we remove the oscillation term caused by the initial population imbalance and, therefore, we can still reliably determine the chiral current.

In a second series of measurements, we investigated the momentum distribution of the system along the $y$ direction after time-of-flight expansion as a function of $K/J$. In the experimentally realized gauge, each quasimomentum $q$ has two real momentum components in the first Brillouin zone located at $k_y=q\pm \pi/(4d_y)$ \cite{Huegel:2013}. Therefore, for the Meissner phase where the lowest energy band has a single ground state at $q=0$, the momentum peaks are located at $k_y=\pm\, \pi/(4d_y)$. In the vortex phase the energy band has two ground states at $\pm \,q_{K/J}$  that depend on the ratio $K/J$, and correspondingly four momentum peaks at $k_y=q_{K/J} \pm\, \pi/(4d_y)$ and $k_y=-q_{K/J} \pm\, \pi/(4d_y)$ are expected (see inset in Fig.~\ref{Fig_5}). When $K/J\ll(K/J)_c$ the two outer peaks at  $k_y=\pm\, q_{K/J} \pm\, \pi/(4d_y)$ vanish and the two inner peaks converge to $k_y=0$.  In order to study this behaviour, we used the same experimental sequence as above, but instead of projecting into isolated double wells along $y$, we directly released the atoms from the trap and determined the time-of-flight momentum distribution. For the Meissner phase, we observe the two expected peaks, but in the vortex phase we only observe the two inner peaks and cannot resolve the position of the outer peaks. The reason for this is that close to the critical point the two peaks at $k_y=\pm\, q_{K/J}+ \pi/(4d_y)$ (and at $k_y=\pm\, q_{K/J} -\pi/(4d_y)$) are too close to each other, and the band flatness combined with the finite temperature do not allow to resolve the two peaks. On the other hand, for $K/J\ll(K/J)_c$ where we could expect to resolve them, the peaks are well separated but the outer peaks vanish. For the analysis of the momentum distributions, we therefore fitted the position of the two inner peaks and measured their relative distance as a function of $K/J$. As can be seen in Fig.~\ref{Fig_5}, we obtain a reasonable agreement with the theoretically calculated peak separation, where the small reduction in amplitude can be explained by considering the finite temperature of the system, which slightly reduces the separation of the momentum peaks as shown by the light green shaded area. There is a two-fold reason for the reduction in the separation due to finite temperature: Non-zero temperature implies population of a fraction of the energy band, which means that due to our experimental gauge the maximum of the peaks are shifted closer to each other \cite{Huegel:2013}. The second reason is that the peaks get broader and are then more strongly affected by the Wannier envelope in the time-of-flight expansion, which also shifts the peaks to a closer position. Additionally, we observed that near the critical point the widths of the fitted peaks have a maximum, which is consistent with the especially flat band at this critical point and with the presence of the outer peaks that cannot be resolved~\cite{Supplements}.\\
\indent In conclusion, the work presented here marks the first demonstration of a low-dimensional Meissner effect and the first observation of a Meissner effect for a bosonic lattice superfluid. It also demonstrates an efficient way to implement spin-orbit coupling in one-dimensional ultracold quantum gases. In future works it would be intriguing to use the recently developed high resolution imaging \cite{Bakr:2010,Sherson:2010} to measure the lattice currents in a {\em spatially resolved way}. This would enable one to not only directly detect the vortices in the flux ladders, but also measure their full current statistics \cite{Kessler:2013}. Measuring the edge currents precisely would also open intriguing avenues for exploring their connection to the edge states of an integer quantum Hall insulator \cite{Huegel:2013}. Furthermore, one could also hope to realize new many-body phenomena in the strongly interacting limit of a Mott insulator \cite{Fisher:1989,Greiner:2002}, where the existence of chiral Mott insulators \cite{Dhar:2012,Zaletel:2013} and a Spin-Meissner effect for two-component systems have been predicted \cite{Petrescu:2013}. Detecting the quantum fluctuations \cite{Endres:2011} of a chiral Mott insulator would enable one to directly probe the chiral currents in this topologically highly non-trivial insulating phase.\\
\indent We thank S.~Nascimb{\`e}ne, Y.-A.~Chen, D. H{\"u}gel and C. Schweizer for their useful comments and for sharing their ideas. This work was supported by the DFG (FOR801), NIM and the EU (UQUAM, SIQS). M. Aidelsburger was additionally supported by the Deutsche Telekom Stiftung.

\section{\si}

\bigskip
\renewcommand{\theequation}{S\arabic{equation}}
\setcounter{equation}{0}
 \renewcommand{\thesection}{S.\Roman{section}}
 
 \section{S1: Experimental sequence}
 \setcounter{figure}{0}
\renewcommand{\thefigure}{S\arabic{figure}}

\begin{figure}[H]
\begin{center}
\includegraphics[width=\columnwidth]{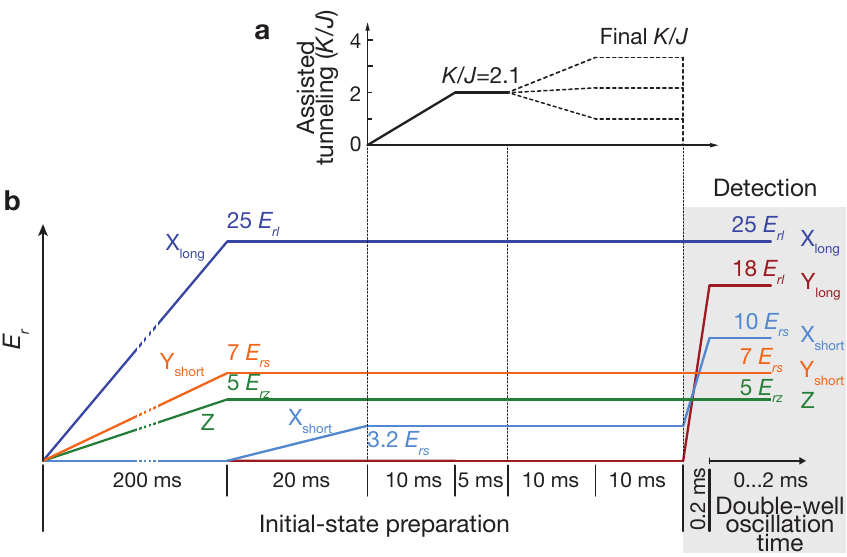}
\caption{Schematic drawing of the experimental sequence for the current measurements showing ramping times and depths of the lattices as well as the strength of the laser-assisted tunneling between the legs of the ladders. In the first part of the sequence, the ground state in the ladders with magnetic flux is prepared for a certain value of $K/J$. This state is then projected into isolated double wells along $y$ to measure the chiral currents.}\label{fig:S1}
\end{center}
\end{figure}

The experimental sequence for measuring the chiral currents described in the main text is shown schematically in Fig.~\ref{fig:S1}. In the first part of the sequence, the system was prepared in the ground state of the ladders in the presence of the magnetic flux. For this, a BEC of about $5 \times 10^4$ atoms was first loaded adiabatically in 200\,ms into a 3D lattice with $V_{lx}=25(1)\,E_{rl}$, $V_y=7.0(2)\,E_{rs}$ and $V_z=5.0(2)\,E_{rz}$, where  $E_{ri}=h^2/(2m\lambda_i^2)$, $i\in \{s,l,z\}$. The tilted ladders were then formed by ramping up the short lattice along $x$ to its final value $V_x=3.2(1)\,E_{rs}$ in 20\,ms, where the phase of the superlattice along $x$ was chosen such that the offset was $\Delta/h=5.57(4)\,$kHz and all atoms ended up on the left side of the ladders. In the next step, laser-assisted tunneling between the initially isolated legs of the ladders was switched on by ramping up the running-wave beams to $V_K^0=4.1(2)\,E_{rK}$ corresponding to $K/J = 2.1(1)$ in 10\,ms, where $E_{rK}=h^2/(2m\lambda_K^2)$ and $\lambda_K=2\lambda_s$. After a holding time of 5\,ms that ensures an equal left-right distribution of atoms (see section S5), the running-wave beams were changed to their final value within 10\,ms. After a subsequent holding time of 10\,ms, we measured the current amplitudes by projecting the system into isolated double well potentials along the $y$ direction. This was done by suddenly switching off the running-wave beams followed by ramping up a long lattice of wavelength $\lambda_l$ along $y$ to $V_{ly} = 18.0(8)\,E_{rl}$ in 0.2\,ms. During this same time, the short lattice along $x$ was increased to $V_x=10.0(3)\, E_{rs}$ to isolate the legs of the ladders. Following a variable holding time between 0 and 2\,ms, where the atoms oscillate in the double wells, the even-odd fraction in each leg was determined using the site-resolved band mapping technique described in section S2.

For the measurements in the ladders with zero flux shown in Fig.~3b and Fig.~4 in the main text, we used the same sequence as described above, but with $\Delta=0$, $V_x=11.0(3)\,E_{rs}$ and without running-wave beams. The resulting ladders had a bare coupling along the rungs of $J_x/h= 250(10)\,$Hz, and the total flux was zero due to the absence of the Peierls phases in the hopping terms. The used projection sequence differs from the one described above in that  $V_x$ was ramped to $20E_{rs}$ instead of $10E_{rs}$ to prevent left-right tunneling. 

\section{S2: Site-resolved detection}

\begin{figure}[h]
\begin{center}
\includegraphics[width=\columnwidth]{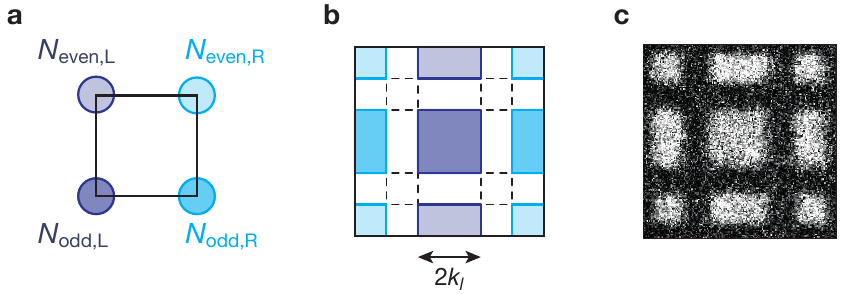}
\caption{\textbf{(a)} Schematic drawing of a four-site square plaquette labeled by the corresponding occupation numbers.
\textbf{(b)} Brillouin zones of the 2D lattice ($k_l=2\pi/\lambda_{l}$).
\textbf{(c)} Typical momentum distribution obtained after the band-mapping sequence, measured after 10\,ms of time-of-flight.}\label{fig:S2}
\end{center}
\end{figure}

After the projection into double wells, the system consists of an array of isolated 2$\times$2 plaquettes with negligible tunneling between neighboring double wells and between the legs of the ladders. The occupation numbers in the four sites of each plaquette on the even/odd and left/right (L/R) locations are denoted by $N_{\textrm{odd};L}$, $N_{\textrm{odd};R}$, $N_{\textrm{even};L}$, and $N_{\textrm{even};R}$. These occupation numbers are extracted by transferring the populations to different Bloch bands, similar to the technique described in refs.~\cite{foelling2007direct,SebbyStrabley2007atomnumber}. A subsequent band mapping allows us to determine the populations in different Bloch bands by counting the corresponding atom numbers. The colors in Fig.~\ref{fig:S2} illustrate the connection between the long-lattice Brillouin zones and the corresponding lattice sites. The evolutions $N_{\textrm{odd};L}(t)$ and $N_{\textrm{even};L}(t)$ ($N_{\textrm{odd};R}(t)$ and $N_{\textrm{even};R}(t)$) are used to evaluate the amplitude of the currents in the left (right) leg of the ladders and $n_{\textrm{odd};\mu}(t)=N_{\textrm{odd};\mu}(t)/(N_{\textrm{even};\mu}(t)+N_{\textrm{odd};\mu}(t))$ with $\mu=L,R$.

\section{S3: Time evolution in double wells}
After preparing the ground state of the flux ladder, we projected the state into isolated double well potentials along the $y$ direction, as described in section S1. Once in the double wells, the atoms oscillate between even and odd sites. The oscillation dynamics on each double well depends on the initially projected state. As described in the main text, the average even-odd oscillation has the form
\begin{align*}
n_{\textrm{even};\mu}(t)-&n_{\textrm{odd};\mu}(t) =\\
&[n_{\textrm{even};\mu}(0)-n_{\textrm{odd};\mu}(0)]\textrm{cos}(2J_\textrm{D}t/\hbar)\\
&-\frac{j_{\mu}}{J/\hbar}\textrm{sin}(2J_\textrm{D}t/\hbar).
\end{align*}

 \begin{figure}
\begin{center}
\includegraphics[scale=1]{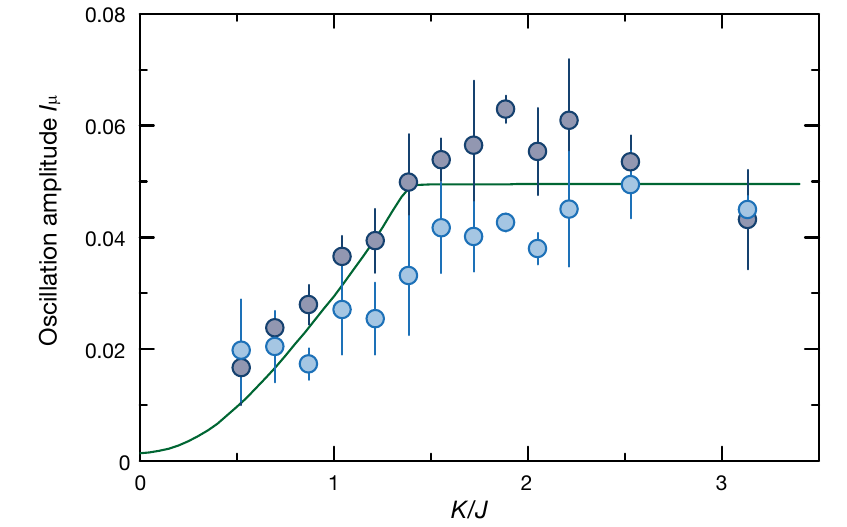}
\caption{Measured oscillation amplitudes $I_\mu$ on the left (dark blue) and right (light blue) legs of the ladder. Each data point is an average over three individual measurements. The error bars denote the standard deviation of the individual measurements. The solid line is the fit of the theoretically calculated chiral current amplitude (divided by two) to our data from Fig.~4 of the main text. The experimental amplitudes are $\sim 14\%$ of the predicted ones for a single non-interacting ladder. Some possible reasons for this reduction in amplitude are the inhomogeneities in the system, imperfect projection of the ladder ground state into the double wells, interactions in the system, heating produced by the running-wave beams, and also the presence of tubes in the transverse ($z$) direction where there is a very weak lattice.}\label{fig:S3}
\end{center}
\end{figure}

 According to the symmetry of the wave function along the ladder, the initial averaged population imbalance should be zero, and the oscillation amplitude should be proportional to the normalized current $j_{\mu}$ (Fig.~\ref{fig:S3}). However, due to experimental imperfections in the system, there might be a small initial population imbalance which leads to an oscillation component that is in phase in the left and right leg (first oscillating term in the expression above). Therefore, when the current amplitude is small, i.e., when $K/J\ll1$, both terms compete and this affects the total oscillation amplitude and phase. However, if the left-right symmetry of the wave function density is preserved, this population imbalance is the same on both sides of the ladder, and they cancel when calculating the chiral current $j_C$. From our data we can obtain an upper bond for the average population imbalance which is less than $\sim20\%$ of the maximum measured oscillation amplitude for all values of $K/J$. We also estimated the amplitude of a possible current running in the same direction on both legs of the ladder which could explain the slightly different amplitudes $I_L$ and $I_R$ for a given value of $K/J$ (see Fig.~\ref{fig:S3}). From our data we conclude that if such current was present, then its amplitude would be smaller than $20\%$ of the maximum measured oscillation amplitude. We note, however, that when calculating $j_C$ this term will also cancel and will not affect our measurements.\\

\begin{figure}
\begin{center}
\includegraphics[scale=1]{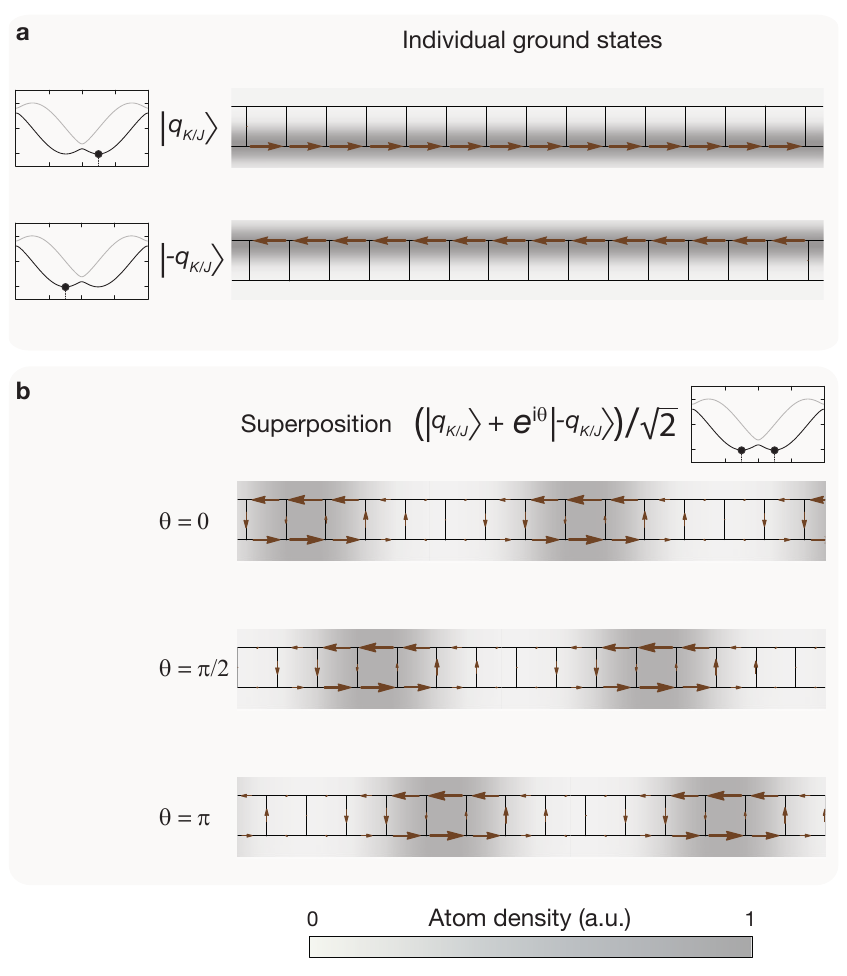}
\caption{Currents and densities for different ground states. \textbf{(a)} Current strength and population density for the individual ground states $| \pm q_{K/J} \rangle$ for $K/J=0.4$. For smaller and smaller values of $K/J$ the population and currents get increasingly concentrated on a single leg of the ladder, right leg for $|q_{K/J} \rangle$ and left  leg for $| -q_{K/J} \rangle$. \textbf{(b)} Current strengths and wave function density for the superposition state $(|q_{K/J} \rangle+e^{i \theta}|-q_{K/J} \rangle)/\sqrt{2}$ calculated for different values of $\theta$. The only effect of $\theta$ is to shift the wavefunction without affecting the average current and density values. As in Fig.~2 in the main text, we have subtracted a constant density offset in the plot and normalized to the maximum density for a better illustration of the spatial density modulation.}\label{fig:S3_2}
\end{center}
\end{figure}

\section{S4: Currents and density for the degenerate ground states}

As shown in Fig.~1 in the main text, in the vortex phase the ladder system exhibits two energy minima at $| \pm q_{K/J} \rangle$ in the lowest band. Each of the two ground states presents different features: $| q_{K/J} \rangle$ ($|-q_{K/J} \rangle$) has most of the population located on the right (left) leg and the current flows mostly on the right (left) leg, with the same chirality in both cases (see Fig.~\ref{fig:S3_2}a). In our experiment, we observe the same population on both legs of the ladder, which is consistent with having populated a superposition state $(|q_{K/J} \rangle+e^{i \theta}|-q_{K/J} \rangle)/\sqrt{2}$ (see next section). This argument is also supported by numerical calculations, which show that for a finite size system with an harmonic trap, the ground state is a linear superposition of $| \pm q_{K/J} \rangle$ with the same weight on both components. A calculation of the currents and density of the system for different values of $\theta$ shows that the phase only shifts the currents and density globally (see Fig.~\ref{fig:S3_2}b).

\section{S5: Left-right distribution}

\begin{figure}[h!]
\begin{center}
\includegraphics[scale=1]{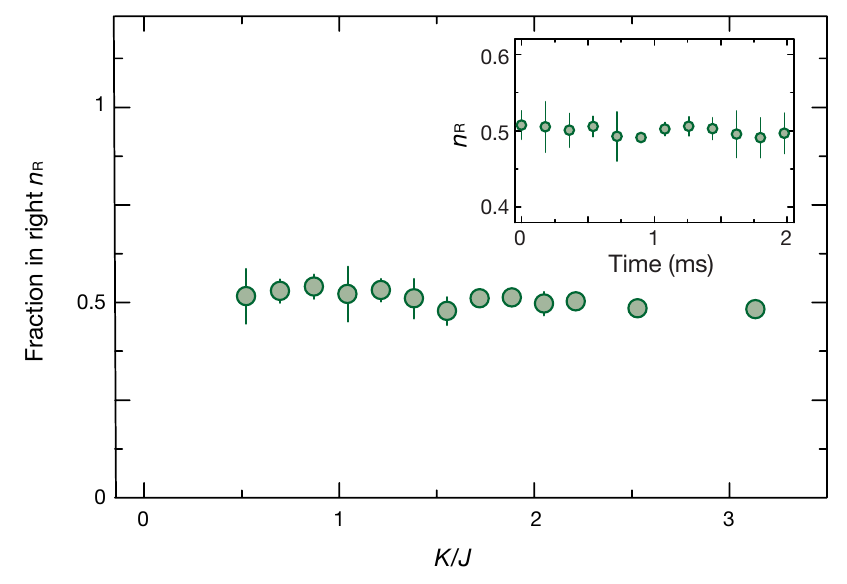}
\caption{Fraction of atoms in the right leg of the ladders for the current measurements shown in Fig.~4 in the main text. All data points are an average over three independent measurements, in which the atom fraction was averaged over the entire evolution and the error bars indicate the standard deviations. The inset shows the time evolution after the projection into the double wells for $K/J = 2.2(1)$, corresponding to the measurements shown in Fig.~3c in the main text. Each point is an average over three individual measurements and the errors are the corresponding standard deviations}\label{fig:S5}
\end{center}
\end{figure}

The site-resolved detection technique, which is used to determine the atom number on the different sites of the plaquettes, allows for a simultaneous measurement of currents as well as atom populations on the left and right legs of the ladders. Figure~\ref{fig:S5} shows the fraction of atoms on the right side for the current measurements of Fig.~4 in the main text, proving that the number of atoms is the same on both sides after the preparation of the final state which started with all atoms initially in the left leg. In addition, no changes in the left-right distribution were observed during the oscillations in the double wells, as expected since both legs are essentially decoupled during this time.

\section{S6: Lifetime of the chiral currents}

\begin{figure}[h]
\begin{center}
\includegraphics[scale=1]{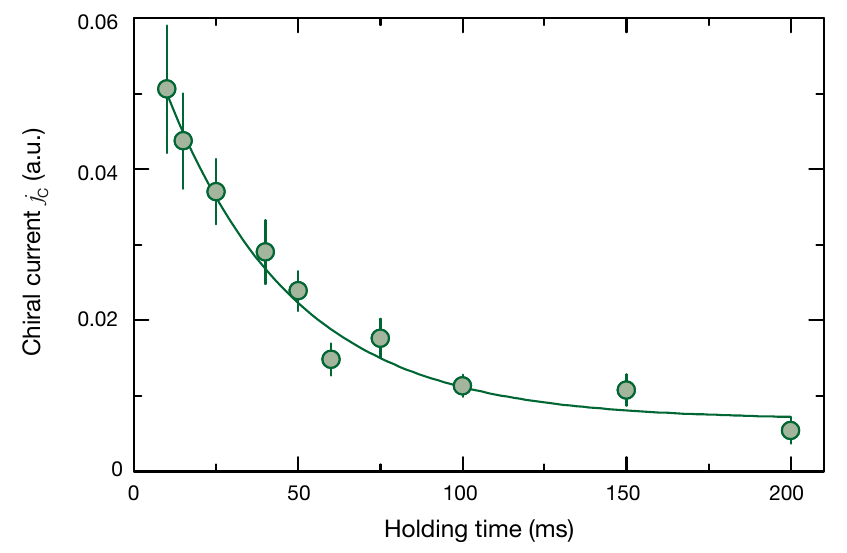}
\caption{Measured chiral current amplitude for $K/J = 2.1(1)$ as a function of holding time before projecting into double-wells. The data points and error bars are determined from three individual measurements as for Fig.~4 in the main text. The solid line is a fit of an exponential decay to the data, which gives a lifetime of 39(7)\,ms.}\label{fig:S4}
\end{center}
\end{figure}

To determine the lifetime of the chiral currents in the ladders, a series of measurements of the current amplitude was performed for different holding times after the preparation of the final state for a ratio of $K/J = 2.1(1)$. Fitting an exponential decay to the data shown in Fig.~\ref{fig:S4} results in a lifetime of 39(7)\,ms. The two main mechanisms leading to the damping of the measured oscillations are most likely heating caused by the running-wave beams as well as decoherence between the individual ladders as the measured amplitude is averaged over the entire system.

\section{S7: Numerical simulations}

\begin{figure}[h]
\begin{center}
\includegraphics[width=8cm]{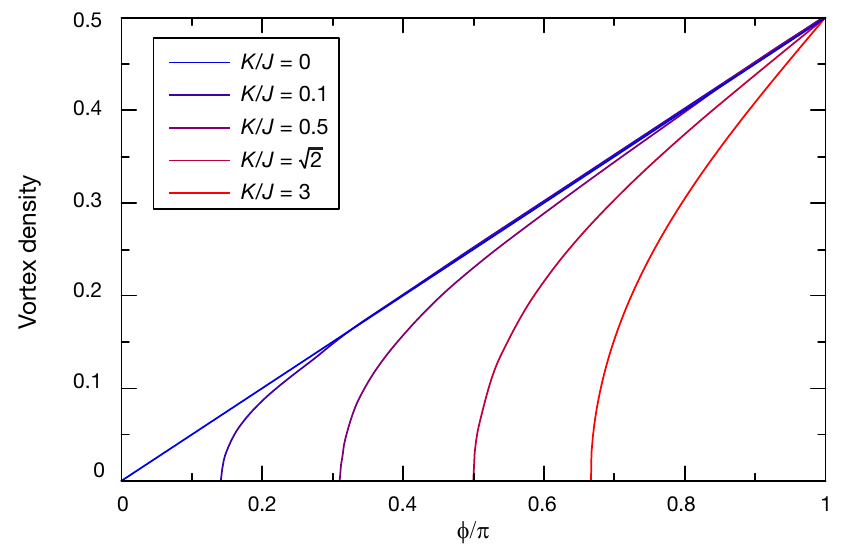}
\caption{Vortex density as a function of the applied magnetic flux for various values of $K/J$ showing the transition from the vortex-free Meissner phase to a vortex phase with finite vortex density for $\phi > \phi_C$.}\label{fig:S6}
\end{center}
\end{figure}

A numerical simulation was used to estimate the effects of the trap as well as the occupation of excited states on the current amplitude. In order to do so, the ground state of a homogeneous ladder with a length of 100 sites was calculated for different values of $K/J$. Following the experimental sequence, this state was then projected onto the two lowest eigenstates of the individual double wells along the ladder. From the time evolution of the state in a double well, one can extract the expected oscillation amplitude by summing over the entire ladder, where the contribution from each double well is weighted according to its occupation probability in the ground state of the ladder. A weak harmonic trapping potential along the ladder with a trapping frequency of $\approx 25$\,Hz present in the experiment causes a slight smoothing of the transition, but the general shape of the curve is not affected. A similar effect can be seen when taking into account a small occupation of excited states.

We also evaluated the density of vortices in the system as a function of the applied magnetic flux. For this, the density distribution of the ground state wavefunction was calculated in a ladder with a length of 300 sites by numerical diagonalization, where we used more sites as above to minimize finite-size effects (this same system was also used to calculate the currents and densities plotted in Fig.~2b in the main text). Changes of the atomic density are directly linked to the vortices in the current and can therefore be used to determine their size.  The results for different values of $K/J$ are shown in Fig.~\ref{fig:S6}, where one can see the transition from the Meissner phase, without any vortices, to the vortex phase when the flux is increased beyond the critical value. For $\phi \gg \phi_C$ the vortex density approaches the value for a decoupled ladder with $K=0$, where the magnetic field penetrates the system completely and the vortex density increases linearly with the applied flux.

\section{S8: Experimental setup, Laser assisted tunneling}

\begin{figure}[h!]
\begin{center}
\includegraphics[width=8cm]{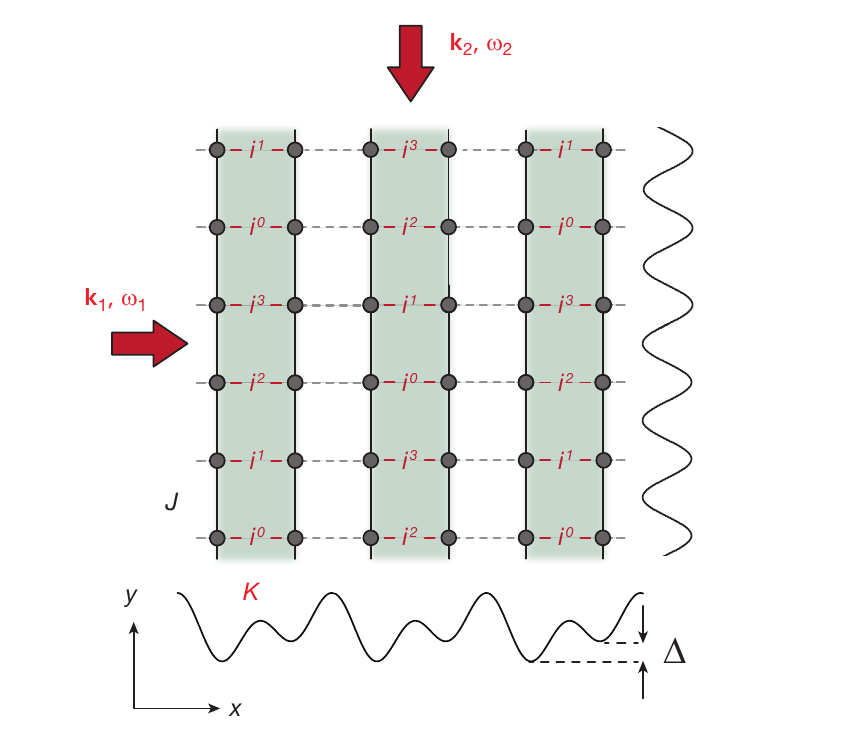}
\caption{Complex spatially dependent hopping terms realized in our experiment. Along the horizontal direction there is a bare tunneling $J_x$ and an energy offset $\Delta$ that inhibits left-right tunneling inside each ladder. The pair of running-wave beams with wavevectors $|\textrm{\textbf{k}}_1|\approx |\textrm{\textbf{k}}_2|=2\pi/\lambda_{K}$ and frequency difference $\omega=\omega_1-\omega_2=\Delta/\hbar$ restores the tunneling and induces a complex hopping term $K$, which has a spatially dependent phase as shown on the rungs of the ladders.}\label{fig:S7}
\end{center}
\end{figure}

Along the $x$ direction, we use a superlattice that creates an array of isolated double well potentials with a coupling $J_x$ and an energy tilt $\Delta$. In the limit $J_x/ \Delta \ll 1$, left-right tunneling is inhibited, and as in our previous experiments we create an artificial magnetic field by using a pair of running-wave beams with frequency difference $\hbar(\omega_2-\omega_1)=\Delta$ that modulate the on-site potential and restore the left-right hopping. The left-right effective coupling has an amplitude $K=J_x \mathcal{J}_1\left(V_K^0/(\sqrt{2}\Delta)\right)\simeq J_xV_K^0/(2\sqrt{2}\Delta)$ and a spatially dependent phase distribution as shown in Fig.~\ref{fig:S7}. The tunneling along $y$ is determined through $J=J_y \mathcal{J}_0\left(V_K^0/(\sqrt{2}\Delta)\right)$. The total phase accumulated by a particle when completing a closed trajectory on a single plaquette is $\phi=\pi/2$, and corresponds to the magnetic flux per unit cell. By changing the wavelength of the running-wave beams or the angle between them, one can in principle engineer any flux. For the measurements shown in Fig.~3 in the main text, we reverse the flux to $\phi=-\pi/2$ by changing the frequency difference to $\hbar(\omega_2-\omega_1)=-\Delta$.

\section{S9: Analysis of momentum distribution in time-of-flight images}

\begin{figure}[h]
\begin{center}
\includegraphics[width=8cm]{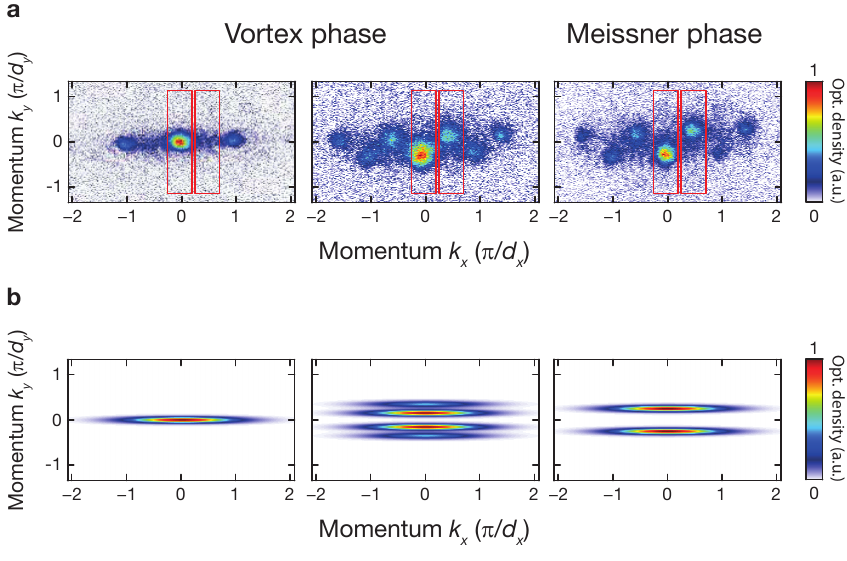}
\caption{Momentum distribution. \textbf{(a)} Time-of-flight images of the momentum distribution and \textbf{(b)} theoretically calculated momentum distributions for $K/J=0.00(1)$ (left), 1.28(5) (center) and 2.1(1) (right). Along the $x$ direction a simple Gaussian was assumed. The densities are normalized to the maximum value for each $K/J$.}\label{fig:S8}
\end{center}
\end{figure}

Figure \ref{fig:S8}b shows the theoretical momentum distribution for three different values of $K/J$, where the ladders are oriented along the vertical ($y$) direction. Along the horizontal ($x$) direction no coherence between the different ladders was assumed. In the Meissner phase there are two momentum peaks in the first Brillouin zone at $k_y=\pm\, \pi/4d_y$. At the bifurcation point the peaks split into four peaks that continue separating, and for $K/J\ll(K/J)_c$ the two outermost peaks vanish and the two inner ones converge to a single peak at $k_y=0$  \cite{Paredes2013}. Fig.~\ref{fig:S8}a shows typical experimental images after $10$\,ms of time-of-flight for the same values of $K/J$, as used for the measurements shown in Fig.~5 of the main text. There, one can see that in the Meissner phase, we can distinguish the two peaks very well. However, due to the finite temperature of the system and the band flatness near the critical point, we cannot distinguish the four peaks in the vortex phase near $K/J=(K/J)_c$, and we can only resolve the two inner ones. For $K/J\ll(K/J)_c$ the four peaks are separated enough to be in principle distinguishable, but since the two outermost peaks vanish we still only observe two inner peaks. In our experiment, we observe also interference effects along the horizontal direction which are due to a residual coherence between individual ladders. 

\begin{figure}
\begin{center}
\includegraphics[width=8cm]{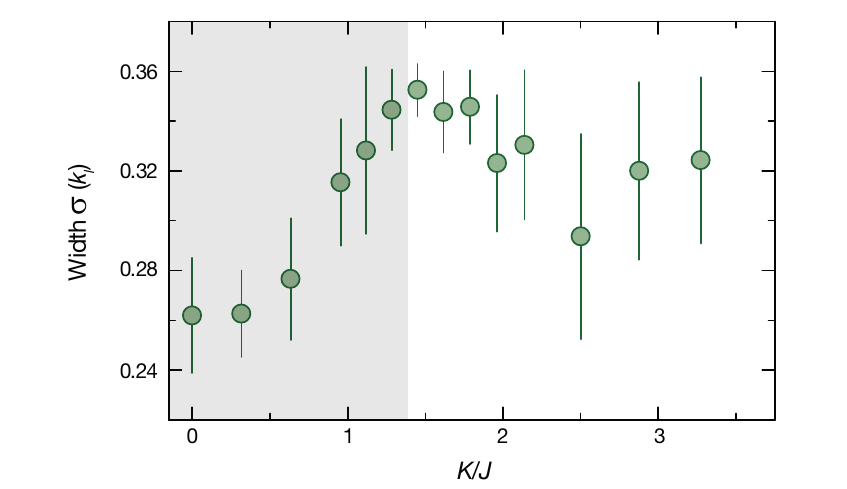}
\caption{Width of the momentum peaks after time-of-flight expansion for the measurements from Fig 5 in the main text. Each point is an average of 5-40 images, and the error bars are the corresponding standard deviations.}\label{fig:S9}
\end{center}
\end{figure}

For the analysis of the peak positions for Fig.~5 in the main text, we first selected two boxes around the two peaks at the center (shown in Fig.~\ref{fig:S8}a) and integrated the signal along the horizontal direction for each box independently. Then we fitted a Gaussian function to each of the two integrated signals and extracted the two peak positions $k_{y1}$ and $k_{y2}$ and the widths $\sigma_1$ and $\sigma_2$. We discarded all the images with a width $\sigma=\sqrt{\sigma_1^2+\sigma_2^2}>0.4 k_y$, for which the peak identification was not reliable anymore, and then calculated the peak separations $k_{y1}-k_{y2}$. Figure \ref{fig:S9} shows the averaged values of the fitted widths  $\sigma$ for the data shown in Fig.~5 in the main text. We observed that it increases with $K/J$ and has a maximum value near the critical point. We attribute this to the presence of the outer peaks at $k_y=\pm\, q_{K/J} \pm \,\pi/(4d_y)$  that have a non-neglible weight near the critical point. Additionally, the energy band becomes very flat near $(K/J)_c$, which combined with the finite temperature of the system increases the width of the peaks.

The light green shaded area shown in Fig.~5 in the main text indicates the theoretically calculated peak separation at finite temperature for the temperature range from 10nK to 30nK. For the calculation, we considered a ladder system with a density of 25 particles per single site of the ladder. A further effect that is also considered in this estimation was the one produced by the Wannier envelope in the time-of-flight expansion. Its effect is to bring the peaks closer, and it becomes more important when the temperature is high, as the width of the peaks gets larger in that case.

\end{document}